\documentclass{piparticle-final}
\usepackage[utf8]{inputenc}
\usepackage{amsmath}
\usepackage{graphicx}

\usepackage{cite} 
\usepackage[normalem]{ulem}  
\usepackage{epstopdf} 

\begin{document}
\volume{7}               
\articlenumber{070014}   
\journalyear{2015}       
\editor{C. A. Condat, G. J. Sibona}   
\received{20 November 2014}     
\accepted{7 September 2015}   
\runningauthor{J. I. Pe\~na Rossell\'o \itshape{et al.}}  
\doi{070014}         

\title{Piezoelectric energy harvesting from colored fat-tailed fluctuations: An electronic analogy}

\author{J. I. Pe\~na Rossell\'o,\cite{inst1} \hspace{0.3em} 
            R. R. Deza,\cite{inst1}\thanks{E-mail: roberto.deza@gmail.com} \hspace{0.3em}
            J. I. Deza,\cite{inst2} \hspace{0.3em}
            H. S. Wio\cite{inst3}  }

\pipabstract{
Aiming to optimize piezoelectric energy harvesting from strongly colored fat-tailed fluctuations, we have recently studied the performance of a monostable inertial device under a noise whose statistics depends on a parameter $q$ (bounded for $q<1$, Gaussian for $q=1$, fat-tailed for $q>1$). We have studied the interplay between the potential shape (interpolating between square-well and harmonic-like behaviors) and the noise's statistics and spectrum, and showed that its output power grows as $q$ increases above 1. We now report a real experiment on an electronic analog of the proposed system, which sheds light on its operating principle.
}

\maketitle

\blfootnote{
\begin{theaffiliation}{99}
   \institution{inst1} IFIMAR (UNMdP-CONICET), De\'an Funes 3350, B7602AYL Mar del Plata, Argentina.
   \institution{inst2} DONLL-UPC, Rambla Sant Nebridi s/n, E-08222 Terrassa, Spain.
   \institution{inst3} IFCA (CSIC-UC), Avda. Los Castros s/n, E-39005 Santander, Spain.
\end{theaffiliation}
}

\section{Introduction}
The race towards downsizing and integration that began half a century ago with the invention of the transistor, necessarily implied developing suitable energy sources for the small, portable device range (which includes wireless sensor systems, self-powered microelectronics, autonomous battery recharging, and many other applications). We have been perhaps too confident on developing better batteries and fuel cells, and have lagged behind on alternatives that would increase the device's autonomy. Among them, extracting useful work out of those motions present in the ambient which we usually regard as ``noise'' or ``fluctuations'' (thermal noise, light, potential differences in living tissue) is an option that has become more and more convenient. In this context, it is possible to conceive devices (meteorological stations, mechanical acquisition or control devices, etc) which---mimicking what living cells do all the time, by absorbing oxygen and nutrients from their surroundings in an autonomous fashion---
could operate indefinitely in their medium without frequent maintenance \cite{1}.

Linear devices are ill-suited to answer this challenge, since they exhibit resonance within a very narrow frequency range. A new and innovative approach---pioneered by Gammaitoni \emph{et al.} \cite{2,2a,3}---calls for nonlinearity as a central ingredient. A nonlinear oscillator can pick energy up from a wider range of frequencies than its linear counterparts, thus better exploiting the available noise energy. Now, much as a linear oscillator needs tuning in order to maximize the energy harvested in a particular situation, a nonlinear one would need to adjust first its potential shape and then its parameters to optimally respond to a noise having certain spectral and statistical features. This work aims thus to optimize the overall performance of a model oscillator, as an energy harvester of supra Gaussian and even L\'evy-like mesoscopic fluctuations through piezoelectric conversion.

Regarding shape, in Ref.\ \cite{4} we have considered a Woods--Saxon \cite{5} oscillator (a monostable potential capable of interpolating between square-well and harmonic-like behaviors). The rationale for this choice is twofold:
\begin{enumerate}
\item On the one hand, a sensible requirement on the abstractions dealt with in the model is that they keep close to what is feasible (in that sense, the polynomial potentials considered in \cite{3} are a too local approximation).
\item On the other hand---as the authors in \cite{3} recognize---, their explicit conclusion (``\ldots\emph{it is possible to demonstrate that nonlinear oscillators can outperform the standard linear ones}.'') is far from being universal and, in fact, a more robust conclusion would be that the harder the polynomial potential, the worse the performance.
\end{enumerate}
Our hypothesis was then that for strongly correlated fat-tail fluctuations, a square well would act as a selector of large excursions, thus enhancing performance. Of course, this would be valid for smoother well potentials, of which the Woods--Saxon one is but one possible family. 

This said, we have performed a thorough numerical study of the dependence of the output rms voltage $V_\mathrm{rms}$ (as a gauge of the mean power delivered to a load circuit) on the potential parameters, as well as on the noise's. We have found a spectacular increase of $V_\mathrm{rms}$ for deep potential wells and low noise intensity, as the noise becomes fat-tailed. Those results led us to conclude that a deep  square-well potential acts as a selector of the large highly correlated oscillator excursions provoked by the fat-tailed noise. In order to further explore that mechanism, we performed a real experiment on an incomplete but illustrative electronic analog: we let noise of the kind considered in Ref.\ \cite{4} pass through a Zener diode, and found indeed that the frequency of Zener current peaks increases as $q$ grows larger than 1.

\section{The model}
Following Ref.\ \cite{2,2a,3}, we consider a one-dimensional anharmonic oscillator which mediates between a source of mechanical vibrations and a piezoelectric transducer producing a voltage \(V(t)=K_c\,x(t)\) to be fed into a load circuit with resistance $R$ and capacitance $\tau_p/R$. In turn, the transducer reacts back on the oscillator with a force \(K_v\,V(t)\).\footnote{Neither delay nor nonlinearity are assumed in the piezoelectric couplings.} The oscillator (of mass $m$ and damping constant $\gamma$) is governed by a monostable potential \(U(x)\), and the source of mechanical vibrations (to which it is coupled with strength $\sigma$) is regarded as stochastic, producing a strongly colored instantaneous force \(\eta(t)\). The system is thus described by
\begin{subequations}\label{eq:1}
\begin{eqnarray}
m\ddot{x}&=&-U'(x)-\gamma\dot{x}-K_vV+\sigma\eta(t),\label{eq:1a}\\
\dot{V}&=&K_c\,\dot{x}-\frac{1}{\tau_p}V.\label{eq:1b}
\end{eqnarray}
\end{subequations}
Being $V^2(t)/R$ the instantaneous power delivered to the load resistance, the measure of perfomance is $V_\mathrm{rms}$ during the observation interval.

\begin{figure}[ht]
\begin{center}
\includegraphics[width=0.9\columnwidth]{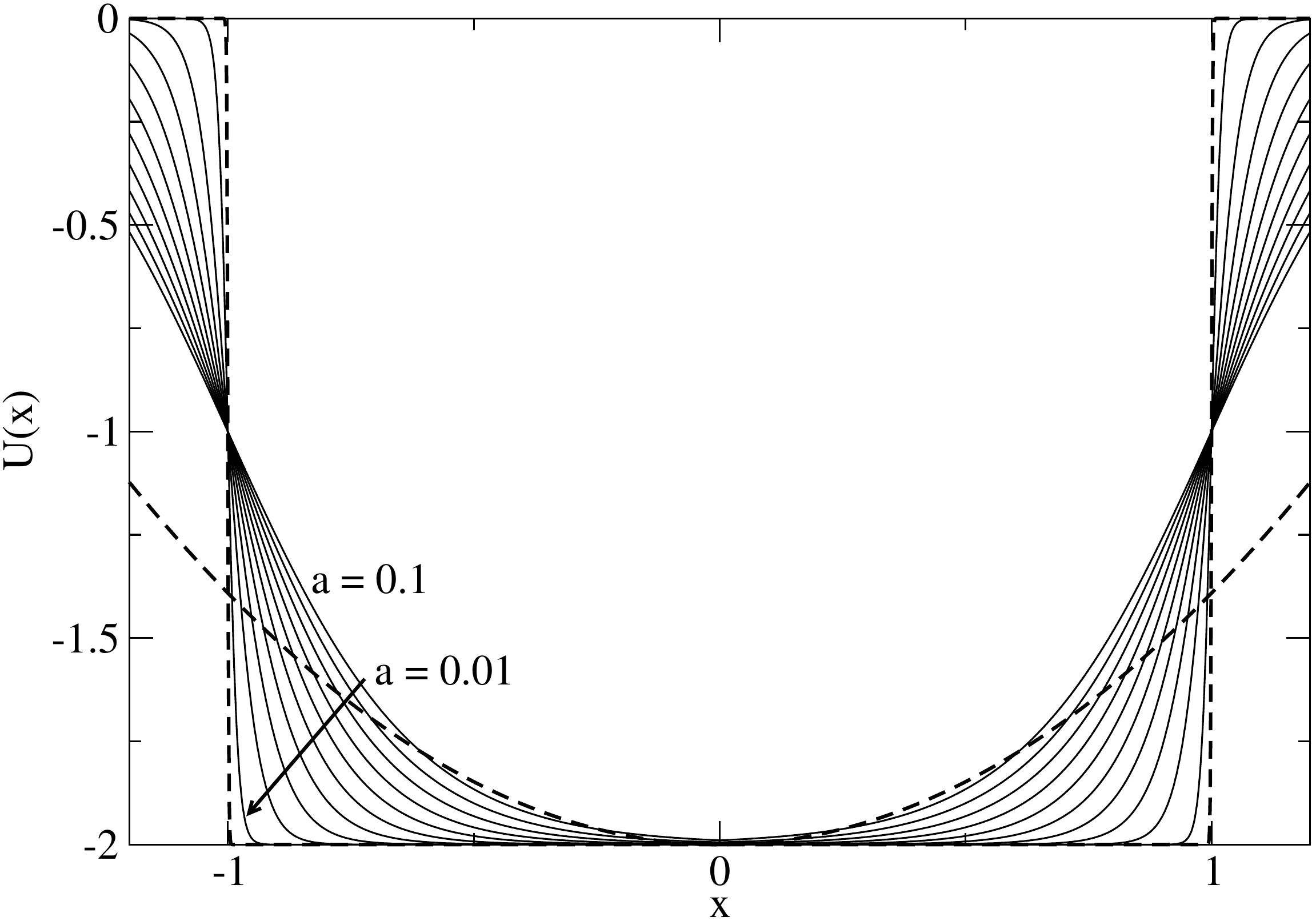}
\end{center}
\caption{WS potential for $r=1$, $V_0=2$ and $a$ between 0.01 and 0.1.}
\label{fig:1}
\end{figure}

We investigate the aptitude of the Woods--Saxon (WS) potential
\begin{equation}\label{eq:2}
U(x):=-\frac{V_0}{1+\exp{\left(\frac{|x|-r}{a}\right)}},
\end{equation}
depicted in Fig.\ \ref{fig:1}, whose behavioral repertoire when varying $a$ spans from that of a square well (with depth $V_0$ and width $2r$) to that of an almost harmonic oscillator. 

The non-Gaussian colored noise $\eta(t)$ chosen in Ref.\ \cite{4} obeys the dynamic equation
\begin{equation}\label{eq:3}
\tau\dot\eta=-V'_q(\eta)+\xi(t),
\end{equation}
with
\begin{equation}\label{eq:4}
V_q(\eta):=\frac{1}{\tau(q-1)}\ln\left[1+ \tau(q-1)\frac{\eta^2}{2}\right],
\end{equation}
where $\xi(t)$ is Gaussian, with $\langle\xi(t)\rangle=0$ and $\langle\xi(t)\xi(t')\rangle=2\delta(t-t')$. $\eta$ is Ornstein--Uhlenbeck for $q=1$, of bounded intensity for $q<1$, supra Gaussian (with variance $2\sigma^2/[\tau(5-3q)]$) for $q>1$ and qualitatively similar to L\'evy's for $q\ge5/3$ (in fact, for $q=2$ it is nonetheless than Cauchy's).

\section{Numerical results}
\begin{figure}[ht]
\begin{center}
\includegraphics[width=0.9\columnwidth]{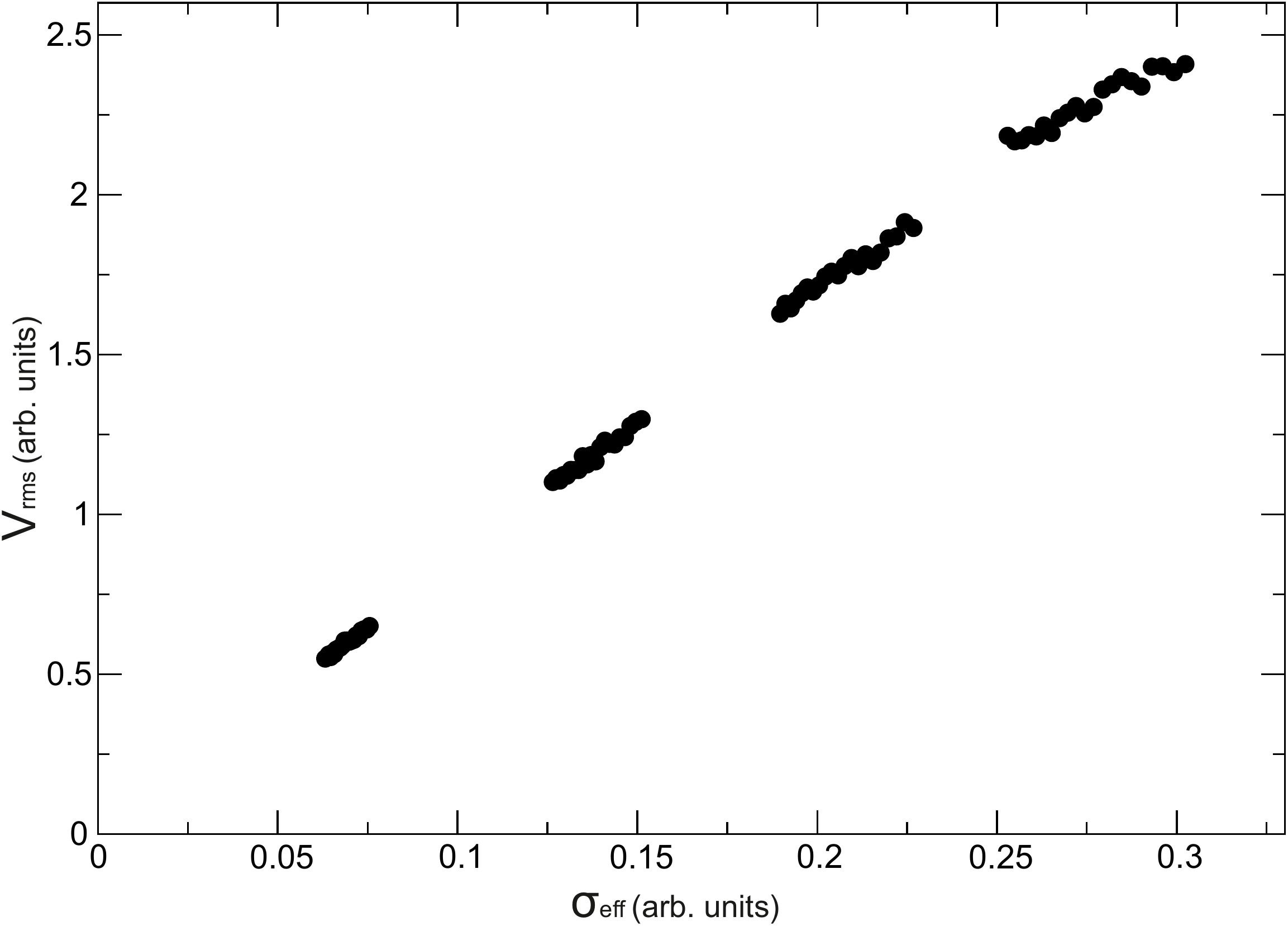}
\end{center}
\caption{\(V_\mathrm{rms}\) \emph{vs.} $\sigma_\mathrm{eff}$ for $V_0=10$, $\sigma=0.2$, and \(a=0.05\).}
\label{fig:2}
\end{figure}

Equations (\ref{eq:1}) and (\ref{eq:3}), together with the definitions (\ref{eq:2}) and (\ref{eq:4}), were integrated in Ref.\ \cite{4} by Heun's method for $m$, $\gamma$, $K_v$, $K_c$ and $r$ set to 1, $\tau_p=10^4$ and $\tau=10$ as in Ref.\ \cite{3}, \(a=0.05\) in Eq.\ (\ref{eq:2}), and several values of $\sigma$ and $V_0$. A sizable increase of $V_\mathrm{rms}$ was found as $q$ in Eq.\ (\ref{eq:4}) rises above 1 for low $\sigma$ and high $V_0$. In fact, $q$ and $\sigma$ are not independent, as shown in Fig.\ \ref{fig:2} by plotting $V_\mathrm{rms}$ \emph{vs.} $\sigma_\mathrm{eff}:=\sigma\sqrt{2/[\tau(5-3q)]}$.

\section{Experiment on an (incomplete) electronic analogy}
\begin{figure}[ht]
\begin{center}
\includegraphics[width=0.9\columnwidth]{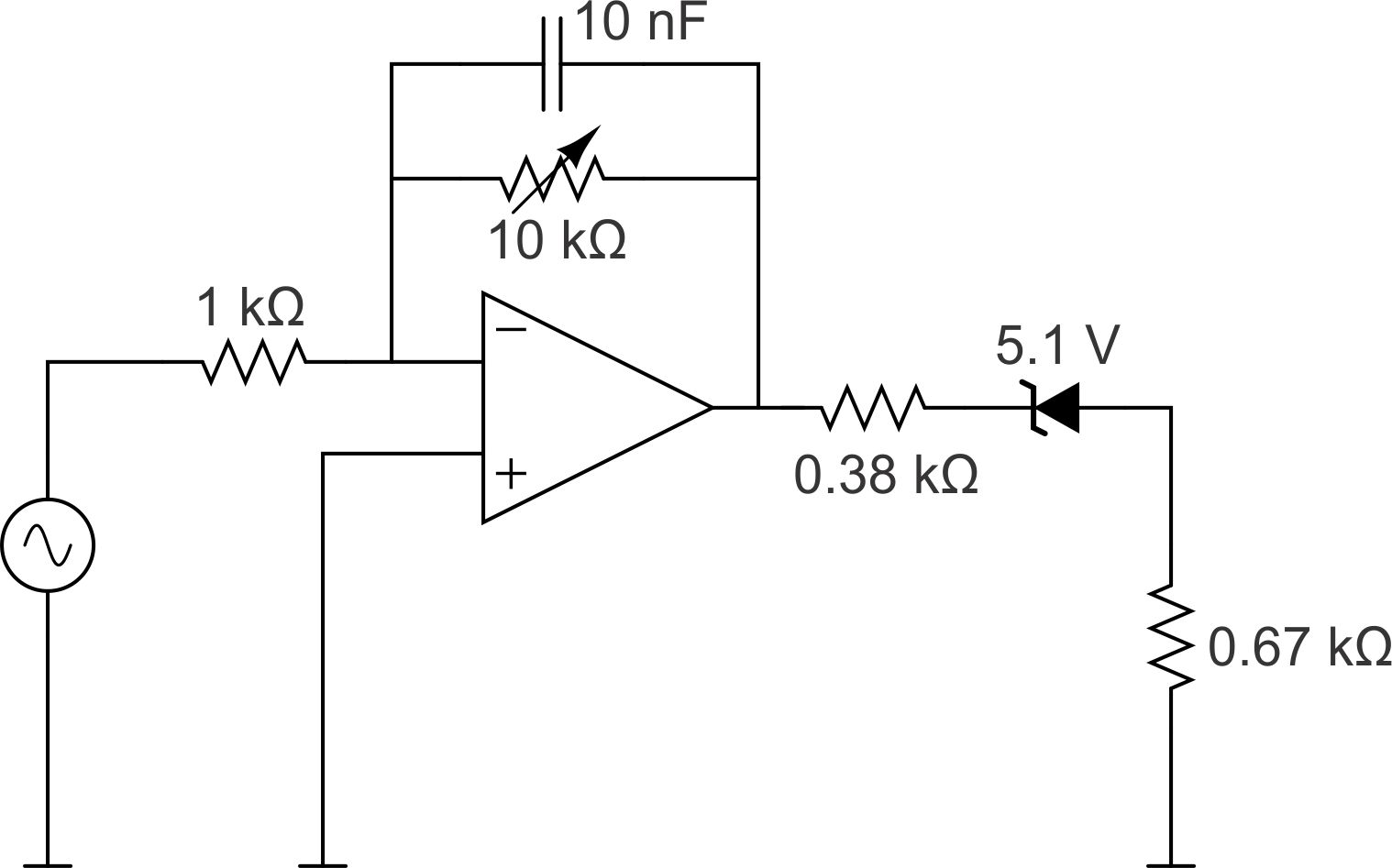}
\end{center}
\caption{Experimental setup. The signal generator is a computer.}
\label{fig:3}
\end{figure}

\begin{figure}[ht]
\begin{center}
\includegraphics[width=0.9\columnwidth]{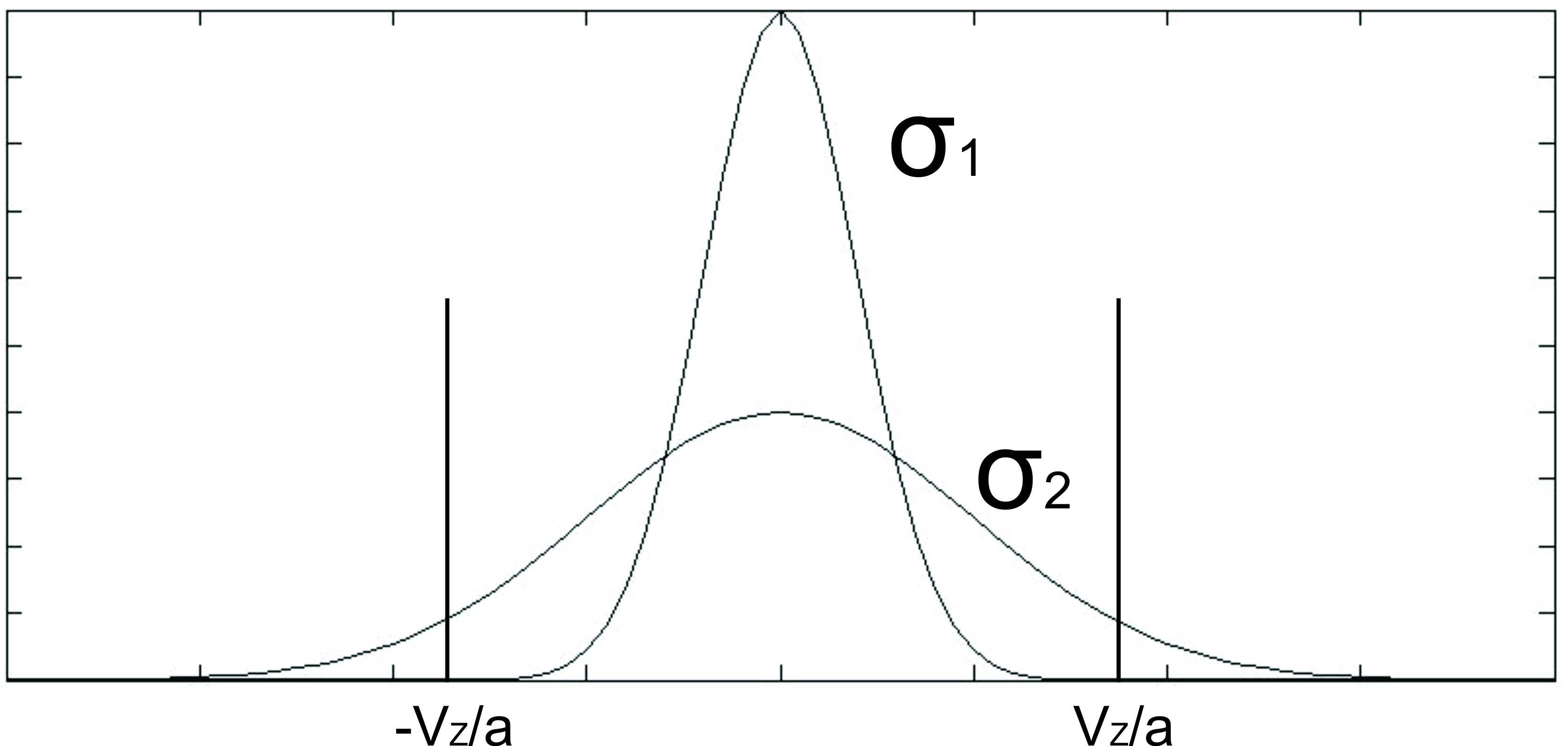}
\end{center}
\caption{Abscissas: sketch of noise voltage at the negative OP AMP input (arb. units). $V_z$ is the Zener voltage and $a$ the feedback OP AMP gain. Ordinates: sketch of the noise PDF for two different effective variances (arb. units).}
\label{fig:4}
\end{figure}

\begin{figure}[ht]
\begin{center}
\includegraphics[width=0.9\columnwidth]{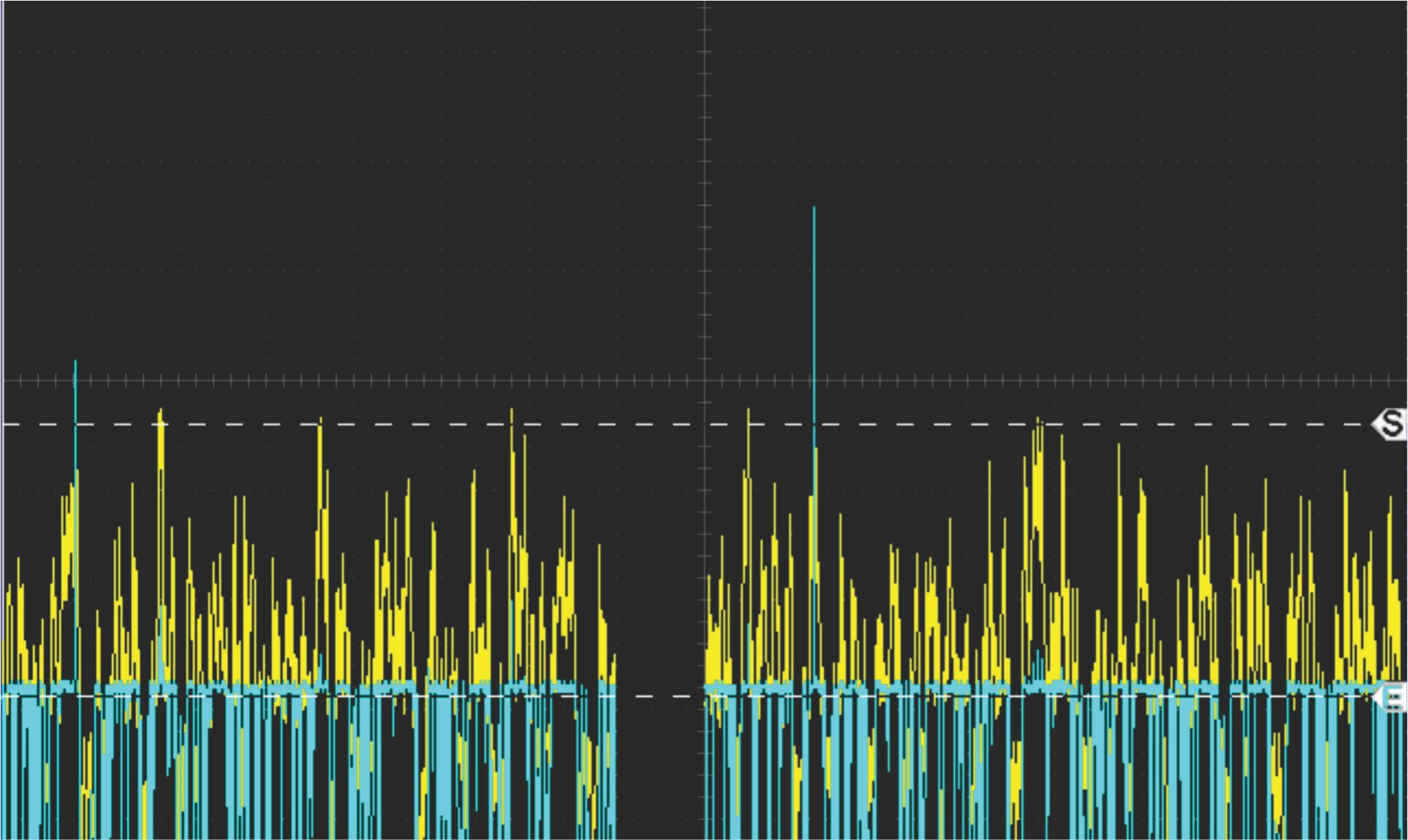}
\end{center}
\caption{Oscilloscope capture for $q=1.6$ and $\sigma=0.4$, showing Zener current peaks (blue) above the dashed line labeled by E, whenever the input noise (yellow) overcomes the bound on Fig. 4 (here represented by the dashed line labeled by S). Abscissas: scope sweeping time (ms). Ordinates: input voltage (V).}
\label{fig:5}
\end{figure}

This growth of $V_\mathrm{rms}$ with $q$ was explained by the potential well's ability to act as a selector of large self-correlated excursions. As a metaphor of this behavior, a Zener diode is used. $\eta(t)$ noise synthetized by means of Eq.\ (\ref{eq:3}) is fed to the circuit in Fig.\ \ref{fig:3} using a \texttt{MATLAB} function, through the computer's audio output. The OP AMP is required because the \texttt{MATLAB} output is limited between $\pm\,2$V. As $\sigma_\mathrm{eff}$ increases (Fig.\ \ref{fig:4}), Zener current peaks can be observed as illustrated in Fig.\ \ref{fig:5} for $\sigma_\mathrm{eff}=0.4$ (outside the range of Fig.\ \ref{fig:2}).

\section{Conclusions}
In a regime where \(V_\mathrm{rms}\propto x_\mathrm{rms}\), it will be dominated by large excursions of the oscillator. These are allowed by a finite-range potential like the Woods--Saxon one, whereas they are suppressed by the even-power potentials considered in Ref.\ \cite{3}. We have chosen as external noise the process defined by Eq.\ (\ref{eq:3}), because it depends on only two parameters with clear interpretation and it is easy to generate dynamically. In such a situation, a moderately round and deep enough square well acts as a great selector of large excursions and is thus able to take the most out of very weak supra Gaussian and even L\'evy-like fluctuations. We have shown that $V_{rms}$ is in fact a function of $\sigma_\mathrm{eff}=\sigma\sqrt{2/[\tau(5-3q)]}$. An illustrative experiment has been performed on the foundation of the system's performance, whose outcome is consistent with our conclusions from numerical work: a Zener diode---regarded here as a metaphor of the well's ability to 
select 
large excursions---produces a current peak whenever such an event occurs.

\begin{acknowledgements}
R.R.D. and H.S.W. acknowledge support from MinCyT of Argentina and MINECO of Spain, through an international collaboration project. 
\end{acknowledgements}

\raggedbottom
\pagebreak

\end{document}